\documentclass[12pt,preprint]{aastex}

\newcommand{\etal}{\mbox{et~al.}}

\newcommand{\Msun}{\mbox{$M_{\odot}$}}

\newcommand{\Lsun}{\mbox{$L_{\odot}$}}
\newcommand{\oii}{\mbox{[O \thinspace II]}~}

\newcommand{\HH}{\mbox{H$_2$}}

\newcommand{\Ha}{\mbox{H$\alpha$} }
\newcommand{\Hb}{\mbox{H$\beta$} }
\newcommand{\Hg}{\mbox{H$\gamma$} }
\newcommand{\Hd}{\mbox{H$\delta$} }

\def\deg      {{\ifmmode^\circ\else$^\circ$\fi}} 

 \shorttitle{G515 Revisited. I. }
 \shortauthors{Liu et al.}
 
\begin{document}
 
 
 \title{G515, Revisited. I. Stellar Populations And Evidence Of Nuclear Activity
 In A Luminous "E+A" Galaxy*}
 

%
%
 \author{ 
Charles T. Liu\altaffilmark{1},
Eric J. Hooper\altaffilmark{2},
Karen O'Neil\altaffilmark{3},
David Thompson\altaffilmark{4},
Marsha Wolf\altaffilmark{2}, and
Thorsten Lisker\altaffilmark{5}
}
 
 
\altaffiltext{$\star$}{Based on data collected at The Arecibo Observatory, part of the National Astronomy and Ionosphere Center, which is operated by Cornell University under a cooperative agreement with the National Science Foundation (NSF);
Kitt Peak National Observatory, National Optical Astronomy Observatory, which is operated by the Association of Universities for Research in Astronomy, Inc.
(AURA) under cooperative agreement with the NSF;  
The MMT Observatory, a joint facility of the Smithsonian Institution and the University of Arizona; and Steward Observatory, operated by the University of Arizona. 
}  
\altaffiltext{1}{Astrophysical Observatory, Department of Engineering Science and Physics, City University of New York, College of Staten Island, 2800 Victory Blvd, Staten Island, NY  10314}
\altaffiltext{2}{Department of Astronomy, University of Wisconsin-Madison, 475 N. Charter Street, Madison, WI 53706}
\altaffiltext{3}{National Radio Astronomy Observatory, P. O. Box 2, Route 28/92, Green Bank, WV 24944-0002}
\altaffiltext{4}{Large Binocular Telescope Observatory, 933 N. Cherry Ave., Tucson, AZ  85721-0065}
 \altaffiltext{5}{Astronomical Institute, Department of Physics and Astronomy, University of Basel, Venusstrasse 7, CH-4102 Binningen, Switzerland}
 
  
\begin{abstract}
 
We present multiwavelength observations of the very luminous "E+A"
galaxy known as G515 (J152426.55+080906.7), including deep $K_{s}$ imaging, spatially resolved \Ha spectroscopy, and radio observations.
The data, together with detailed spectral synthesis of the galaxy's integrated stellar population, show that G515 is  a $\sim1$ Gyr old post-merger, 
post-starburst galaxy.  We detect no Balmer line emission
in the galaxy, although there is a small amount of [N II] $\lambda\lambda$ 6548, 6583 \AA~ emission.  
The galaxy's H I mass has a 2$\sigma$ upper limit of $1.0 \times 10^{9}$ \Msun.  IRAS detections in the $60\mu$m and $100\mu$m bands indicate a far infrared luminosity of $\sim5.8 \times 10^{10}$ \Lsun.  A small amount ($\sim$ 3 mJy) of radio continuum flux, which appears to be variable, has been detected.  The data suggest that G515 may have once been an ultraluminous infrared galaxy, and may harbor a weak, dust-obscured active nucleus.

\end{abstract}
 
 
 

 
 \section{Introduction}
 
The term "E+A galaxy" was first used to describe galaxies
whose optical spectra in the approximate rest-frame range
of $\sim$ 3600 \AA~-- 5100 \AA~ showed little or no line emission,
deep higher-order Balmer line absorption, and continuum resembling
that of A stars superimposed upon an early-type galaxy spectrum.  
After \citet{dre83} identified E+A galaxies
in distant rich clusters of galaxies, a number of studies have
been conducted, in a variety of galaxy environments and redshift 
ranges, to understand the evolutionary histories of these galaxies,
and possibly their significance as diagnostics of galaxy evolution
both in the field and in clusters.  Such studies have included 
optical spectroscopic and narrow-band imaging surveys in rich clusters of galaxies at high and low redshift (\citealt{cou87,lav88,fab91,cal93,bel95,cal96,cal97,fis98,mor98,dre99}) and in the field (\citealt{zab96,got03,qui04}), optical spectrophotometry (\citealt{liu96})
and stellar population modeling (\citealt{new90,cha94,abr96,bar96,liu96}), and
near-infrared (\citealt{gal00,bal05}) and 
radio (\citealt{sma99,cha01,mil01,got04})
surveys.  From another direction, studies of nearby merging galaxies with post-starburst spectroscopic characteristics (\citealt{sch82,liu95a,liu95b}; Hibbard \& van Gorkom 1996; \citealt{sch96,lai03}) and spatially resolved studies of nearby E+A galaxies (\citealt{nor01,yam05,yag06a,yag06b}) have provided insight into the mechanisms that might produce galaxies with E+A spectra. 

One particularly intriguing example of the E+A galaxy phenomenon was first identified by Oegerle, Hill, \& Hoessel (1991).  Given the arbitrary catalog name G515, it appears to be an example of the end stage of the galaxy-galaxy major merger sequence.   A very luminous E+A galaxy at redshift $z = 0.0875$, it has no detectable \oii emission and deep Balmer absorption lines (the mean equivalent width of \Hb, \Hg, and \Hd absorption = 9.1 \AA).  Its central surface brightness profile follows a $r^{1/4}$ de Vaucouleurs law, but it is asymmetric, including a faint, comma-like tidal tail extending toward the south.  In this first of a series of papers examining this enigmatic system, we assemble data from the literature together with new observations to study its stellar population and star formation history.  We explore the hypothesis of \citet{oeg91} that G515 is a former ultraluminous infrared galaxy, and we also present measurements of radio continuum variability that we interpret as evidence of a weak active nucleus in the galaxy's core.

Throughout this paper, we adopt a flat cosmology
with $\Omega_{\Lambda} = 0.7$, $\Omega_{m} = 0.3$, and
$H_{o} = 70$ km s$^{-1}$  Mpc$^{-1}$.

\section{Data and Observations}

\subsection{Multiband Photometry}

A substantial amount of information about G515 has been obtained in numerous observational programs over the years.  The galaxy is catalogued in the Sloan Digital Sky Survey (SDSS; \citealt{ade06}) as SDSS J152426.50$+$080908.0 and in the Two Micron All Sky Survey (2MASS; \citealt{skr06}) as 2MASX J15242648+0809082.  Aperture photometry in the SDSS $ugriz$ and 2MASS $JHK_{s}$ passbands, measured within apertures of radii $6\arcsec.3$ and $5\arcsec.6$ respectively, are given in Table \ref{phot_1}.  These magnitudes are not corrected for Galactic extinction,
which was estimated at 0.04 magnitude in $r$ by \citet{oeg91}.

As a check to these data, we note that \citet{oeg91} computed an $r-$band isophotal magnitude of 15.24 at $\mu = 24.5$ mag arcsec$^{-2}$.  Using the SDSS $r-$band image, we made that same measurement and 
found  $r = 15.25 \pm 0.03$.  In addition, we compared the 2MASS photometry to that of \citet{gal00}, who measured $(J, H, K_{s}) = (13.86 \pm 0.02, 13.11 \pm 0.05, 12.75 \pm 0.05)$ using slightly smaller apertures.  Correcting for that aperture difference, those measurements are consistent within the published error bars of the 2MASS data.  

Based on these measurements and the galaxy's redshift, G515 has an absolute magnitude of $M_{r} = -22.7$ and $M_{K_s} = -25.4$, making it one of the most luminous E+A galaxies yet cataloged (\citealt{got05}).

\subsection{Near-Infrared and Optical Imaging}

We obtained deep $K_{s}$ imaging of G515 with FLAMINGOS on the 4-meter Mayall Telescope at Kitt Peak National Observatory on the nights of February 6-8, 2004.  A total of ($139 \times 45$ s) exposures were obtained, 
totaling $6255$ s of integration.  The exposures were each dithered 
over several arcminutes in random directions, to move the extended light of the galaxy around on the detector.  In addition to maximizing the accuracy of flat fielding and removing the effects of bad pixels, this also allowed us to use the data to self-sky-subtract without biasing measurements of the faint extended light. 

The photometry was calibrated on the 2MASS scale directly from 68 unsaturated point sources on the image.  The rms scatter of the photometric zeropoint, as fit to these unsaturated sources, was a somewhat high 0.11 magnitude.  As a check, we measured the $K_{s}$ apparent magnitude of G515 against the 2MASS photometry in the same band, using an identical aperture; our result was $K_{s} = 12.69$, which is within the $1\sigma$ error of the 2MASS result.

We present the $K_{s}$ image in the lower right panel of Figure \ref{oir}.  In the other three panels of Figure \ref{oir}, we present imaging data from SDSS: the composite $gri$ ("finder chart") image and the $g$ and $r$ images respectively.  All four panels show, as expected, the same basic morphology: a bright ellipsoidal main galaxy "core", an asymmetric "crescent" extending southwestward from the core, and a faint tidal tail shaped like a backward "comma" extending some $40\arcsec$ southward of the core.  Extending the aperture of G515 beyond the core, from $\sim 5\arcsec$ to $13\arcsec$ to include the crescent, the galaxy's total magnitude increases to $K_{s} = 12.50$.  (A faint, red star 
approximately 10$\arcsec$ to the southwest of the core, most clearly visible in the $K_{s}$ image,
was deblended and subtracted from the galaxy's light in each passband.)
The faintness and extent of the tidal tail makes its photometric measurement more uncertain; within a rectangular aperture $25\arcsec \times 12\arcsec$ aligned with the tidal tail (roughly northeast to southwest), centered $32\arcsec$ south of the center of the galaxy core, its magnitude is
$K_{s} = 17.13 \pm 0.16$.
 
\subsection{Integrated Spectrophotometry}

Integrated spectrophotometry was previously published by \citet{liu96}, and we present it again here for further analysis using current stellar population synthesis models.  We obtained these data  using the Red Channel Spectrograph at the Multiple Mirror Telescope in June 1993, using a 300 lines $mm^{-1}$ grating for the wavelength region $3500 - 6000$ \AA~ and a 270 lines $mm^{-1}$ grating for the wavelength region $5500 - 9500$ \AA. Exposure times were $600$ s for each grating setting.  Our aperture was a $5\arcsec \times 180\arcsec$ slit, and was centered on the bright core of G515.  The large slit width allowed us to obtain integrated spectra at $27$ \AA~ resolution of the objects and avoid problems with differential atmospheric refraction.  

Standard data reduction techniques using the IRAF software system were used to process the long-slit data, primarily with the CCDRED and LONGSLIT packages. The aperture along the slit was traced and extracted with the APEXTRACT package; the spectrum was then deredshifted using the NEWREDSHIFT task in IRAF, rebinned to a rest wavelength dispersion of $10$ \AA~ per pixel, and normalized to unity at $4500$ \AA~ rest wavelength. Particular care was taken to remove saturated major night sky features (e.g., those at 5577 \AA, 5889 \AA, and the ÔÔBÕÕ and ÔÔAÕÕ bands at 6900 \AA~ and 7600 \AA) from the spectra as accurately as possible. 

The results of the new stellar population analysis of this integrated spectrophotometric data 
are presented in Section 3.2.

\subsection{Spatially Resolved Spectroscopy}

Complementing the integrated spectrophotometry of the main core of G515, we obtained higher-resolution longslit spectra of the galaxy and its tidal features on March 6, 1996, using the B\&C Spectrograph on the Steward Observatory 2.3-meter telescope.  We used an 832 lines $mm^{-1}$ grating, together with an $1.5\arcsec \times 180\arcsec$ slit, to obtain $\sim 3$ \AA~ spectral resolution (1.4 \AA~ pixel$^{-1}$) in the wavelength range 6650\AA--7450\AA.  The goal of these observations was to resolve the H$\alpha + $[N II] emission-line complex, at the redshift of G515, into its three components (H$\alpha~ \lambda 6563$ \AA~ and [N II] $\lambda\lambda 6548, 6583$ \AA).

In all, we observed this galaxy at several different slit positions.  We defer the full discussion of all these data to a future paper (C. T. Liu \etal, in preparation), where we will examine more closely the two-dimensional spatial properties of the G515 system.  In this work, we will focus on the slit position which runs through the nucleus of the galaxy. Figure \ref{slit} shows the slit orientation with respect to the galaxy as a whole.  The integration time for this particular spectrum was 2 exposures of 1200 seconds each.  We processed and reduced the spectrum using the standard procedures in IRAF, taking particular care to ensure that wavelength calibration was optimized in the region of observed \Ha $+$ [N II] to an rms error of $< 0.5$\AA.  The spatial scale of the longslit spectrum was $0\arcsec.866$/pixel.

In Figure \ref{spec}, we present the results in the spectral range 7070 -- 7200 \AA, corresponding to a rest wavelength of 6500 -- 6620 \AA, along the spectrograph slit at $0\arcsec.866$ intervals.  Since the atmospheric seeing for these observations was approximately $1\arcsec.4$, this tight spacing means that each spectrum is not quite spatially distinct from one another.  Even with this oversampling, however, it is clear that the [N II] emission doublet appears clearly within several arcseconds of the galaxy core's central peak, with a combined rest-frame equivalent width of a few \AA~ up to $\sim 10$ \AA~ at the peak itself.  On the other hand, neither \Ha emission nor absorption appears, except possibly at  the $\sim1\sigma$ level at the very core, where a small emission line may be emerging from what could be an \Ha absorption trough.  But even in that case, the equivalent width uncorrected for Balmer absorption is less than 1 \AA.

\subsection{Far-Infrared and CO Data}

We examined the far-infrared emission of G515 using the SCANPI utility available via the NASA/IPAC Infrared Science Archive 
(http://scanpi.ipac.caltech.edu:9000/).  The noise-weighted mean scan results are presented in Figure \ref{iras}.  In summary, IRAS detected no flux from G515 in its $12\mu$m and $25\mu$m bands, to $2\sigma$ limits of 0.050 Jy and 0.056 Jy respectively.  Flux was detected, however, in the other two bands: $0.15 \pm 0.019$ Jy at $60\mu$m, and $0.35 \pm 0.051$ Jy at $100\mu$m.  According to the SCANPI output, the signal to noise ratio of these two detections were 8.6 and 8.7 respectively.  

These detections yield a far infrared color of $\log(S_{60}/S_{100}) = -0.37$ for G515.  Following the formulae used in \citet{rie86} and \cite{kim95}, its far infrared luminosity is $5.3 \times 10^{10} \Lsun$ in the $10 - 200\mu$m range, and $5.8 \times 10^{10}$ \Lsun in the $8 - 1000\mu$m range.  These values are comparable to those typically found in the lower-end range of luminous infrared galaxies (see, e.g., \citealt{kim95}).

G515 was also observed by Lo, Chen, \& Ho (1999) in CO$(1\rightarrow0)$ with the NRAO 12-meter telescope at Kitt Peak.  They did not detect molecular gas, 
and they set an upper limit of the \HH~ mass of $7 \times 10^{10} ~ \Msun$.

\subsection{Radio Data}

\subsubsection{H I Observations}

We observed G515 over a nine month period extending from
June 8, 2000 through Febuary 13, 2001, using the refurbished Arecibo 305m telescope.  The data were taken using the L-narrow receiver with 9-level sampling and 2048 lags/polarization. As four independent boards were available, two (one/polarization) were set to a 12.5 MHz bandwidth and two to 6.25MHz, for a raw (unsmoothed) resolution of 1.523 and 0.7615 km/s at 1306.1 MHz (the observed center frequency). Standard position switching techniques were used, and blank sky data were obtained after every five minutes of on-source observation.

The data were calibrated using standard telescope gain curves, obtained within 2-3 months of each observing run.  Telescope temperature was determined by observing calibrator diodes after every 10 minutes of observation.  Details on the L-narrow receiver gain curves and diode measurements can be found at http://www.naic.edu, and calibration information can be found in \citet{one04}.

After calibration, individual 5 minute on-source observations were examined, and any observations which contained excessive noise due to either RFI or sunrise/sunset effects on the telescope were removed from the final data set, leaving a total of 355 minutes of useful on-source observation.

No object was detected, to the limits listed in Table \ref{arecibo_1} below.
The flux and mass upper limits were derived assuming a
2$\sigma$ detection and assuming a 80 km/s velocity width.

\subsubsection{Radio Continuum}

We examined archival 21 cm radio continuum maps produced by the NRAO VLA Sky Survey (NVSS; \citealt{con98}) and the FIRST Survey (\citealt{bec95}) at the position of G515 and its surrounding vicinity of sky.  In Figure \ref{radio}, we present the contour plot of the NVSS field and the radio image of the FIRST field.  

The NVSS data, taken on February 27, 1995, showed a detection at the location of G515, within a position error of $\sim 9\arcsec$, of a 1.4 GHz continuum radio source with a total NVSS catalog flux of $3.0 \pm 0.5$ mJy.  In the FIRST archive, however, the data were obtained on January 25, 2000, and no object was detected.  With a listed point source rms of 0.155 mJy, a constant 3.0 mJy source should easily have been detected.

To investigate whether the NVSS detection might have been a spurious anomaly, we further analyzed the NVSS image from the archive.  We employed a simple point source detection and noise measuring routine ({\citealt{vis92,hoo95}), without any of the small NVSS corrections such as clean bias.  This yielded a flux of 2.9 mJy at the location of G515; and in three patches of "blank" sky near that position, the rms of the background flux was $\sim 0.6 - 0.7$ mJy.  This suggests that the G515 detection is real, at least at the $4\sigma - 5\sigma$ level.  We also checked the larger vicinity on the NVSS map.  There is a 55 mJy source $3\arcmin$ from G515; no other sources greater than 50 mJy are located within $0.45\deg$.  Furthermore, no sources $> 100$ mJy are within $0.75\deg$; and the brightest source within a $1\deg$ radius is a 288 mJy source $0.98\deg$ away.  All of these detections are well within the dynamic range of the NVSS, so it is unlikely that the G515 detection might have been spuriously produced by a strong source nearby.

As a final check, we further analyzed the FIRST radio image in the vicinity of G515.  We made multiple runs near the optical position, and found that the background rms was 0.19 mJy.  Within an $18\arcsec$ radius of the optical position, there is a maximum flux of 0.23 mJy; as this is only $\sim 1.3\sigma$, we confirm that there is no detection of G515 in the FIRST image.

It is possible that the NVSS detection is an extended signal, produced by obscured star formation, which is resolved out of the FIRST image.  Alternatively, the FIRST non-detection may be evidence of variability in the radio continuum emission, the source of which is a weak, obscured AGN in the core of G515.  We discuss these two possibilities in more detail below,
in Sections 3.1 and 3.3 respectively. 

\section{Discussion}
\subsection{Ongoing star formation}

Our H I upper limit of $\sim 1.0 \times 10^9$ \Msun~ of atomic hydrogen, and the molecular gas upper limit from \citet{lo99}, suggests that ongoing star formation is not being strongly fueled by unprocessed or infalling cold gas, if indeed any is occurring at all.  The spectrum of \citet{oeg91} showed no [O II] emission, and the integrated spectrum of \citet{liu96} shows no [O II] or \Ha emission.   The longslit spectroscopy of the \Ha$+$[N II] spectral region shows no \Ha emission detected above the $1\sigma$ level along the length of the slit.  These results appear to confirm the absence of massive stars that produce substantial ionizing UV flux.

All of our data are consistent with the picture that G515 has little or no ongoing star formation.
One possible exception to this interpretation is that, since the higher order Balmer lines have a mean absorption equivalent width of $9.1$ \AA, there could be approximately this amount 
(i.e. $9-10$ \AA~ equivalent width) of \Ha emission that has filled in a deep \Ha absorption trough.  Another possible exception is the NVSS radio continuum detection; heavily dust-obscured star formation could be generating an extended signal, which could in turn have been resolved out of the FIRST image.  The optical spectrum, however, does not exhibit the substantial reddening that might be expected for a large amount of extinction due to dust.

\subsection{Stellar Population}

\citet{liu96} modeled the integrated stellar population of G515 using a simple two-component model: a simple solar metallicity starburst superimposed on an early-type galaxy spectral energy distribution.  They found that one of the best-fitting models was a pure starburst 1.02 Gyr old; good fits were obtained across the rest wavelength range $3200 - 8000$\AA~ with bursts ranging from 0.64 to 1.14 Gyr, and starburst mass fractions ranging from 19\% to 100\%.

Using those same data, we modeled the stellar population again, this time with the updated stellar population synthesis models of \citet{bru03}.  Two significant improvements of these models compared with the GISSEL models used by \citet{liu96} are significantly higher spectral resolution ($3$\AA~ vs. $20-40$\AA) and superior accounting for metallicity as a free parameter.  Because of the low resolution of our integrated spectrophotometry, we obtain relatively little benefit from the models' higher resolution.  The metallicity modeling, however, is an informative new constraint in the modeling.

The model grid includes stellar ages ranging from 0.1 to 14 Gyr 
and metallicities ranging from [Fe/H] of -1.0 to +0.5. Model 
spectra are smoothed to a resolution of 27\AA~ to match the G515 
spectrum and fit over the rest wavelength range of 2590\AA~ to 8400\AA. 
We perform the fits in two ways: a full spectrophotometric fit, 
and a continuum-normalized fit where the continuum shape is removed 
and just the spectral lines are fit (as described in \citealt{wol07}). 
For each technique, we used either a single simple starburst population 
or a two-component fit, and we either allowed the metallicity to 
vary as a free parameter or fixed the metallicity near solar at 
[Fe/H] = -0.1, 0.0, or +0.1.

The results of the model fits are presented in Figures \ref{starf} and \ref{starc}.  It should be noted that the models are generally bluer than the galaxy spectrum; this has been attributed in part to the contribution of thermally pulsing asymptotic giant branch stars (\citealt{mar05}).  Also, the metallicity determinations of these model fits are best interpreted as a relative diagnostic, rather than an absolute value determination.  

With these caveats in mind, the fits show that the full spectrum 
of G515 is indeed best fit by a single stellar population 1.0-1.5 Gyr 
in age, or in the case of a two-component fit, an older (5-14 Gyr) 
population plus a 0.4 Gyr old starburst contributing some 5-10\% 
of the optical light. The stellar populations are similar in the 
continuum-normalized best fits: 0.8-1.5 Gyr for a single burst 
population, or a 5-14 Gyr older population plus a 0.1-0.8 Gyr younger 
population contributing 5-20\% of the luminosity.  The lowest 
$\chi_{r}^{2}$=4.25 value occurs for the two-component 
continuum-normalized fit with metallicity as a free parameter. 
However, the single stellar population fit with the same parameters 
is nearly indistinguishable with $\chi_{r}^{2}$=4.35. The $\chi_{r}^{2}$ 
values for the full spectrum fits are, as might be expected, higher at 6.28-6.75 for the 
two-component fits and 10.05-12.57 for the single stellar population 
fits. Using $\chi_{r}^{2}$ as an indicator of goodness of fit, the 
continuum-normalized models provide the best fits and a single 
stellar population at 1.5 Gyr is indistinguishable from a two-
component fit with an older population at 10-14 Gyr and a younger 
population at 0.8 Gyr contributing 20\% of the light. Very young, 
massive stars are not required to produce the optical spectrum in 
any of these cases.

The broad-band colors in G515 are consistent with these modeling results,
which suggest a blue stellar population indicating a significant, recently completed starburst.
According to the SDSS data, within a $6\arcsec.3$ radius, the integrated observed colors are ($g-r, r-i$) $=$ ($0.54, 0.29$).  In the SDSS photometric system, these are between the colors expected of an Sbc and an Scd galaxy at z=0.088 (see, e.g., \citealt{cap07}), as would be expected for either a single,  decaying starburst or a composite burst-plus-older population.

The observed $r-K_{s}$ colors of the different component parts of the G515 system differ somewhat.  Within the core ($i.e.$ the $6\arcsec.3$ radius), in the extended crescent, and in the tidal tail, $r-K_{s} = 2.77 \pm 0.04, 2.56 \pm 0.06,$ and $2.77 \pm 0.20$ respectively.  (As described in Section 2.2, the faint red foreground star $\sim 10\arcsec$ southwest of the core, near the edge of the crescent, was deblended and removed before measuring the galaxy's magnitudes.) No color term or K-correction has been attempted for these colors, given the low redshift and the uncertainty of the spectral energy distribution; the relative differences, however, between the core (and tail, though the photometric error is high) and the crescent suggest the presence of either somewhat distinct stellar populations, or perhaps inhomogeneous internal dust extinction.  

\subsection{A Hidden AGN in G515?}

E+A galaxies are strongly associated with galaxy interactions and mergers.  Many of the low-redshift E+A galaxies that have been imaged (\citealt{zab96,yam05,yag06a,yag06b}), as well as galaxies with E+A-like, deep Balmer absorption and weak [O II] emission (\citealt{liu95a,liu95b}), either show the morphological signs of past interaction events or are clearly in the midst of a galaxy interaction.  G515, with its asymmetric crescent and its long tidal tail, is no exception to this trend; and it makes sense that a galaxy interaction or merger event triggered the global starburst that has since led to the distinctive E+A spectral signature.

Another important class of galaxies often associated with galaxy interactions are ultra-luminous infrared galaxies (ULIGs), objects whose far-infrared luminosity ($\sim 10-1000\mu$m) is of order $10^{11} - 10^{12}$ \Lsun~ and dominates their overall energy output.   ULIGs generally show asymmetries and tidal features as well, indicating merger activity or interaction, and have a dust-enshrouded nuclear region which harbor a powerful nuclear starburst -- and often, and may always, also harbor an active nucleus (\citealt{san88}).  

If, as the prevailing model suggests, merger-created ULIGs harbor dust-enshrouded active nuclei, then at least some fraction of those central AGN could be formed during the merger, and still remain obscured $\sim 1$ Gyr after the merger-induced starburst.  This may be due to a relatively slow flow of infalling material to feed the accretion-powered emission of the central black hole.  Or, some other dynamical conditions of the merger/interaction could slow or nearly end the AGN duty cycle before the shroud becomes optically thin.  Such an AGN might be expected to produce detectable nonthermal emission but little optical line emission, while the galaxy as a whole would exhibit an E+A spectrum.    If a fraction of such ULIGs have an extended post-starburst phase after their initial nuclear starbursts, then it would be possible to have a luminous, merger-created E+A galaxy with infrared and radio characteristics of a luminous infrared galaxy.

Progenitors of such "post-ULIG E+A galaxies" may have already been observed.  Liu \& Kennicutt (1995b) showed that many ULIGs exhibit E+A-like spectral features, albeit usually with weak rather than no [O II] line emission (a characteristic later dubbed "e(a)";  \citealt{pog00}).  \citet{got06} observed that for some AGN, the extended post-starburst regions of their host galaxies are sharply centered around the nucleus.  Also, \citet{yan06} have suggested that most post-starburst galaxies may harbor AGN or low-ionization, narrow emission-line regions (LINERs).  It is yet unclear, however, if the post-starburst activity can outlast its associated AGN activity.  \citet{car88} showed that about 10\% of shell elliptical galaxies show E+A-like spectra, but it is not known if those galaxies harbor supermassive black holes.  Searches for a central compact object in galaxy mergers have not yet yielded a concrete detection (e.g., \citealt{lai03}).

We have presented evidence that G515 is a post-ULIG E+A galaxy harboring a weak AGN at its core.  \citet{oeg91} suggested this possibility based on G515's optical luminosity, because assuming a merger-induced global starburst has passively evolved for $\sim10^9$ years, its luminosity just as the starburst ended would have been of order $10^{12}$ \Lsun.  The strong current FIR emission of G515, $\log(L_{FIR}) = 10.76$, is further indication of a ULIG origin for G515, though the lack of detection of $12\mu$m and $25\mu$m emission suggests that the dust in the galaxy is not quite as warm as in typical ULIGs.

The lack of optical emission lines, unfortunately, makes the standard identification of an AGN via emission-line diagnostics rather problematic.  The only statistically significant emission lines in either the spatially resolved or the integrated spectra are the doublet [N II] lines.  The width of the stronger line, [N II] $\lambda 6583$ \AA, is $8.7 - 10.2$ \AA~ in the $\sim 3\arcsec$ diameter region at the central peak of the galaxy core; this corresponds to a rest-frame velocity dispersion of $360-430$ km/s, which is near the top of the range for the central velocity dispersions of typical large elliptical galaxies.  The existence of [N II] emission might suggest the existence of a LINER at the nucleus of G515;  but even if that were so, it is not definitive confirmation of the existence of an AGN, since LINER emission can be produced by shock ionization as well as photoionization (\citealt{vei95}).

The radio continuum detection and its variability, as indicated by the NVSS and FIRST survey data, is thus the strongest evidence for an active galactic nucleus at the core of G515.    As discussed earlier, there is no evidence for any spatially extended star formation, so it is unlikely that the NVSS detection is a spatially diffuse signal that has been resolved out of the higher resolution FIRST data.  Since there appear to be no young, massive, hot stars in the galaxy, AGN heating may in fact be producing the FIR luminosity, at this late stage in the ULIG's evolution.  

Our interpretation of the FIRST non-detection as radio variability is statistically strongly justified, but has the caveat that we are using data from two very different surveys as two-epoch data for the same source.  \citet{wan06} successfully used NVSS and FIRST in this way, but only for positive detections in both bands.  Another issue to consider is the implied strength of the radio continuum emission and its variability.  The radio properties of low luminosity AGN span a broad range (\citealt{nag05,fil06}), and G515 falls well within that range;  and variability at the $\sim3$ mJy level is strong but certainly not extreme.  In fact, \citet{nag02} found that nearly half of all local low luminosity AGN have significant inter-year variability.   Future, deeper observations with high spatial resolution will help clarify the source of the radio continuum emission in G515.
 
\section{Conclusions}

In this first of a series of papers, we have confirmed that G515 is a luminous E+A galaxy with no nebular Balmer line emission, indicating no current star formation.  Its optical and near-infrared properties match that of a $\sim 1$ Gyr old stellar population, which was almost certainly created when a galaxy merger or interaction event triggered a large starburst in the galaxy.  Its far-infrared and radio properties suggest that G515 was formerly an ultraluminous infrared galaxy that has slowly faded, and that it may host a weak AGN at its core.  G515 appears to be an example of the end stage of the galaxy-galaxy major merger sequence -- the transition point from a former ULIG to a future giant elliptical galaxy.

 
\acknowledgments

We are grateful to the creators, keepers, and supporters of the the NRAO VLA Sky Survey (available at http://www.cv.nrao.edu/nvss/) and the FIRST survey (available at http://sundog.stsci.edu).
C. Liu acknowledges the hospitality and support of the Hayden Planetarium 
and Department of Astrophysics at the American Museum of Natural History.  
T. Lisker acknowledges support by the Swiss National Science Foundation.
This work was supported in part by a grant from the City University of New York PSC-CUNY Research Award Program (\#68660-00 37) and in part by NASA grant HST-GO-09822.40-A to the College of Staten Island.  We thank the referee for a detailed and constructive review that improved the quality of this paper.  FLAMINGOS was designed and constructed by the IR instrumentation group (PI: R. Elston) at the University of Florida, Department of Astronomy, with support from NSF grant AST97-31180 and Kitt Peak National Observatory.  This publication makes use of data products from the Two Micron All Sky Survey, which is a joint project of the University of Massachusetts and the Infrared Processing and Analysis Center/California Institute of Technology, funded by the National Aeronautics and Space Administration (NASA) and the NSF.  This research has made use of the NASA/ IPAC Extragalactic Database (NED) and Infrared Science Archive (IRSA), which are operated by the Jet Propulsion Laboratory, California Institute of Technology, under contract with NASA.
Funding for the SDSS has been provided by the Alfred P. Sloan Foundation, the Participating Institutions, the National Aeronautics and Space Administration, the National Science Foundation, the U.S. Department of Energy, the Japanese Monbukagakusho, and the Max Planck Society. The SDSS Web site is http://www.sdss.org/.  The SDSS is managed by the Astrophysical Research Consortium (ARC) for the Participating Institutions. 

 
 {\it Facilities:} \facility{Arecibo}, \facility{IRAS}, \facility{KPNO}, \facility{MMT}, \facility{Steward Observatory}, \facility{VLA}.
 
 

 \clearpage
 
 


 \clearpage

 \begin{deluxetable}{ccc}
 \tabletypesize{\scriptsize}
 \tablecaption{Photometry of G515}
 \tablewidth{0pt}
 \tablehead{
 \colhead{Passband} &  \colhead{Magnitude} & \colhead{Source\tablenotemark{a}}
 } 

\startdata

 $u$	& $17.54 \pm 0.013 $ & SDSS		\\
 $g$	& $15.96 \pm 0.003 $ & SDSS		\\
$r$		& $15.42 \pm 0.003 $ & SDSS		\\
$ i$ 	& $15.13 \pm 0.003 $ & SDSS		\\
 $z$	& $14.91 \pm 0.004 $ & SDSS		\\
 $J$	& $13.74 \pm 0.031 $ & 2MASS	\\
 $H$	& $13.06 \pm 0.037 $ & 2MASS	\\
 $K_{s}$	& $12.67 \pm 0.048 $ & 2MASS	\\

 \enddata
 \tablenotetext{a}{SDSS = Sloan Digital Sky Survey; 2MASS = Two Micron All Sky Survey}
 
 \label{phot_1}
 \end{deluxetable}

\clearpage
 
 \begin{deluxetable}{ccccc}
 \tabletypesize{\scriptsize}
 \tablecaption{H I measurements}
 \tablewidth{0pt}
\tablehead{
\colhead{Resolution} &  \colhead{Smoothing}   &  \colhead{RMS}  & \colhead{Flux} & \colhead{Mass} \\
\colhead{(km/s)}  & \colhead{(\# channels)} & \colhead{(mJy)} & \colhead{(Jy km/s)} &  \colhead{(\Msun)}
}

\startdata
 
$1.523$ &   0         &      0.3998  &    $< 0.06396$ &   $< 2.12 \times 10^9$ \\
$15.23$ &   10       &      0.2136  &    $< 0.03418$ &   $< 1.13 \times 10^9$ \\
$30.46$ &    20      &     0.1979   &    $< 0.03170$ &   $< 1.04 \times 10^9$ \\

\enddata
\label{arecibo_1}
\end{deluxetable}

\clearpage
\begin{figure}
\epsscale{1.0}
\plotone{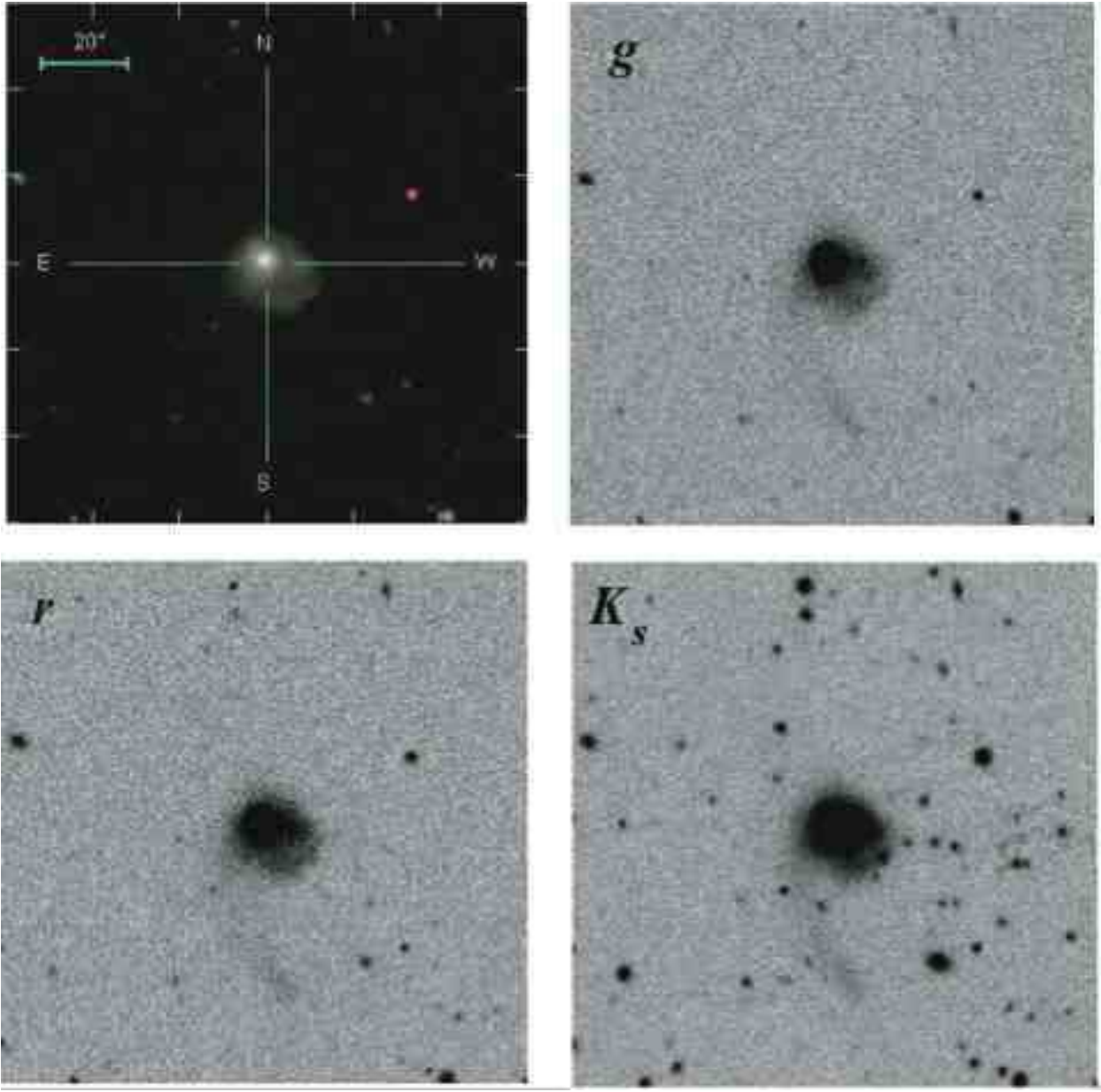}
\caption{
Optical and near-infrared images of G515.  Each panel is 120$\arcsec$ across, and north is up and east to the left for each panel. {\it Upper left:} SDSS $gri$ composite ("finder chart") image. {\it Upper right, lower left:} SDSS $g$ and $r$ images.  {\it Lower right:} $K_{s}$ image obtained with the KPNO 4-meter telescope.
}
\label{oir}
\end{figure}

\clearpage
\begin{figure}[ht]
\epsscale{1.0} 
\plotone{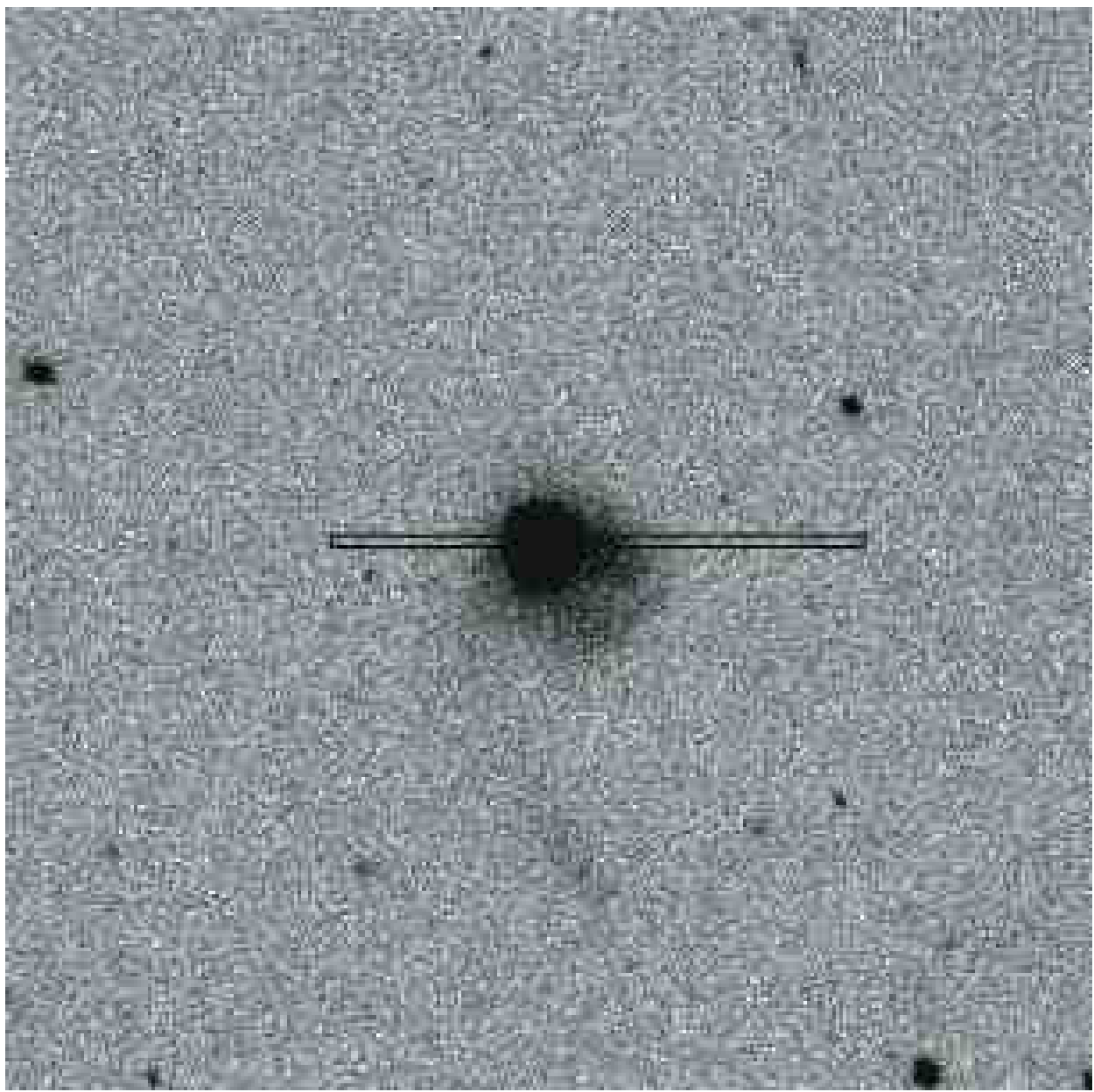}
\caption{
Slit position of the longslit spectroscopy presented in Figure \ref{spec} (see text).
} 
\label{slit}
\end{figure}

\clearpage
\begin{figure}[ht]
\epsscale{1.0} 
\plotone{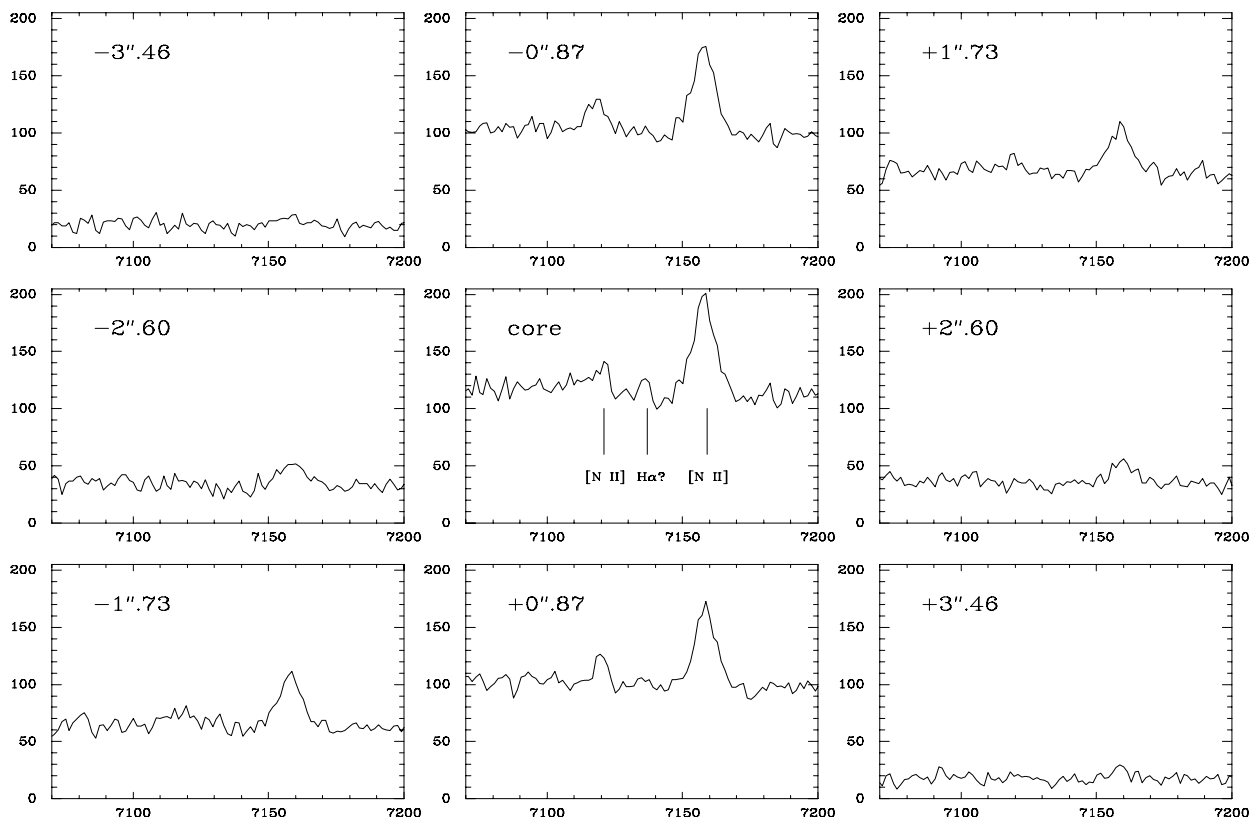}
\caption{
Spectra extracted along the longslit spectrum as shown in Figure \ref{slit}.  Down the columns from left to right, the spectra are $-3\arcsec.46, -2\arcsec.60, -1\arcsec.73, -0\arcsec.87, 0, +0\arcsec.87, +1\arcsec.73, +2\arcsec.60,$ and $+3\arcsec.46$ away from the central core of G515, from west to east.  The \Ha line in the core of G515 would appear redshifted ($z = 0.0875$) to $\lambda 7137$\AA, while the [N II] doublet appears at $\lambda\lambda 7121, 7159$\AA, as shown in the center panel.
} 
\label{spec}
\end{figure}

\clearpage
\begin{figure}[ht]
\epsscale{1.0} 
\plotone{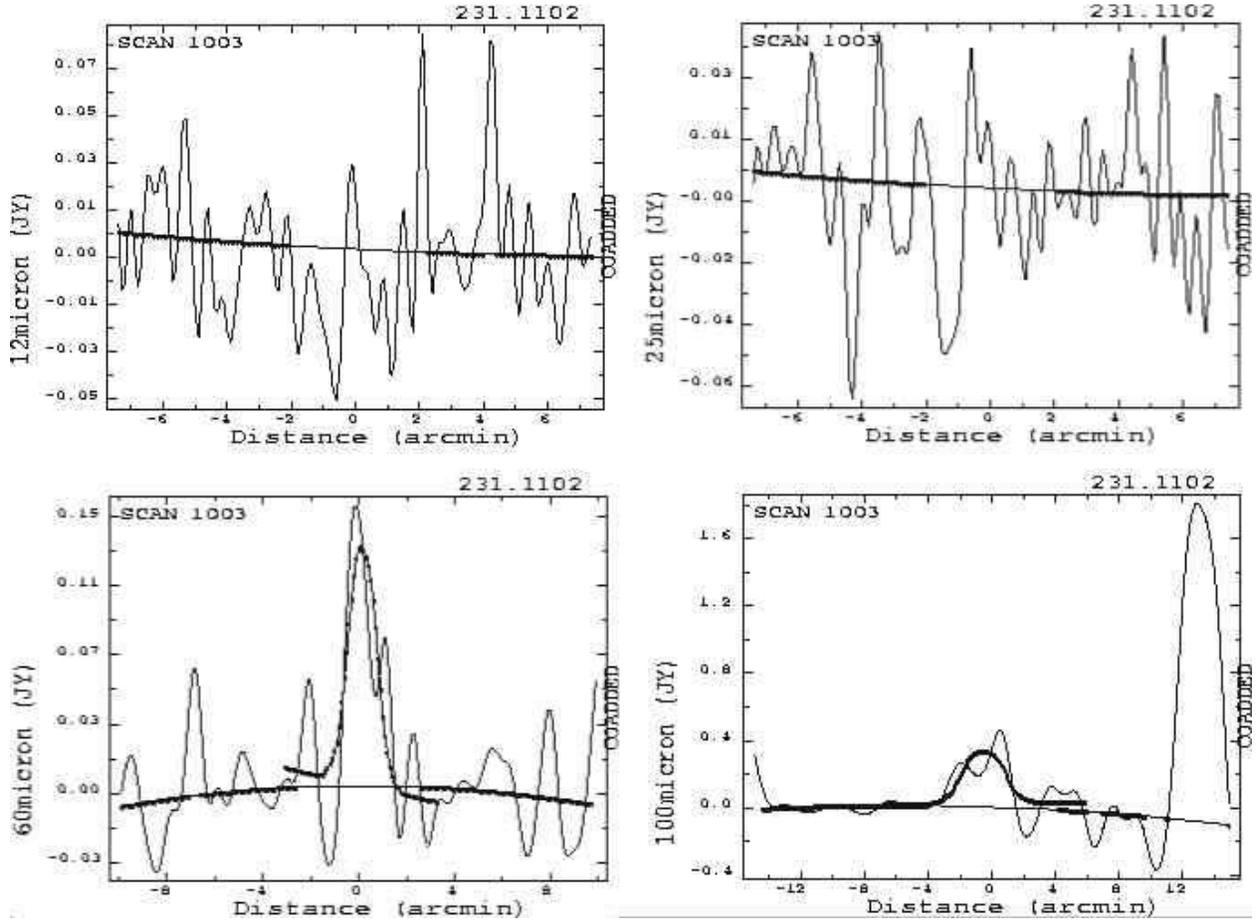}
\caption{
Noise-weighted mean SCANPI output of IRAS observations at the location of G515 (see text for description).  The number "231.1102" at the upper right of each panel is the J2000.0 right ascension of G515 in decimal degrees.  The dark solid line in each panel is the best fit to the IRAS data as computed by SCANPI.  Detections at $8.6\sigma$ and $8.7\sigma$ are apparent in the $60\mu$m and $100\mu$m bands respectively.  The strong detection near the far end of the $100\mu$m scan is of CGCG 077-119, a bright barred spiral galaxy at $z = 0.036.$
} 
\label{iras}
\end{figure}

\clearpage
\begin{figure}
\epsscale{1.0}
\plotone{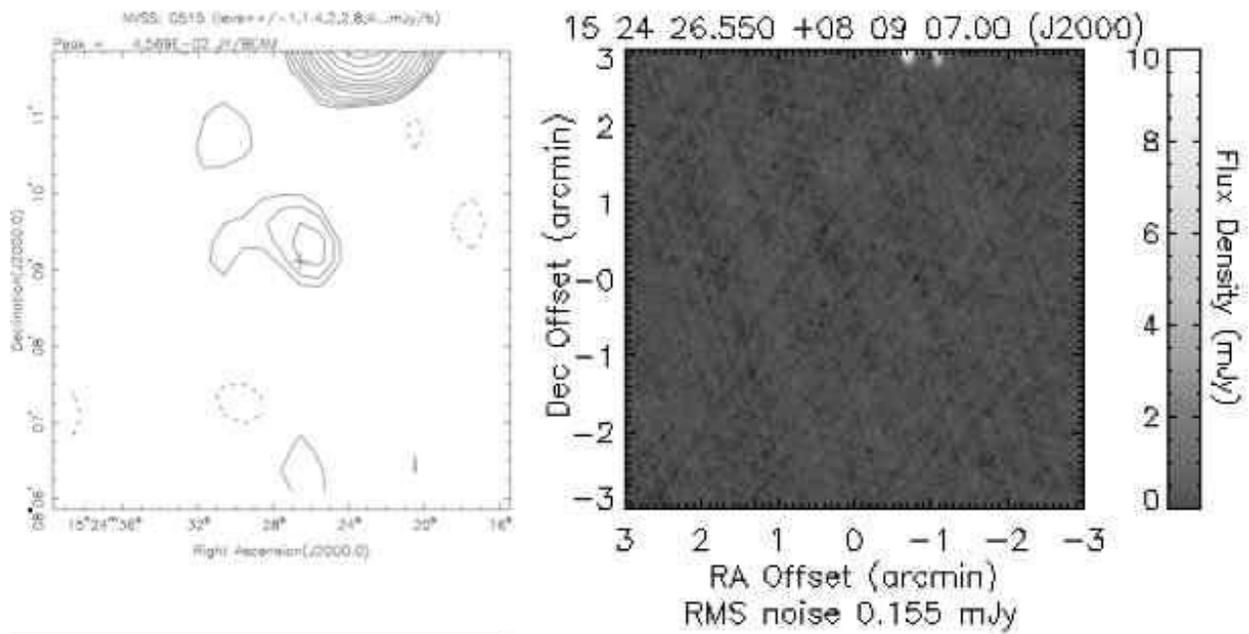}
\caption{
1.4 GHz Radio continuum maps at the position of G515. $Left:$ Contour map from the NVSS archive, observed 27 February 1995.  $Right:$ Image from the FIRST archive, observed 25 January 2000.  Both images are 6$\arcmin$ across, and detect the bright ($\sim 55$ mJy) source at the top edge of the field of view.}
\label{radio}
\end{figure}

\clearpage
\begin{figure}[ht]
\epsscale{1.0} 
\includegraphics[scale=0.40]{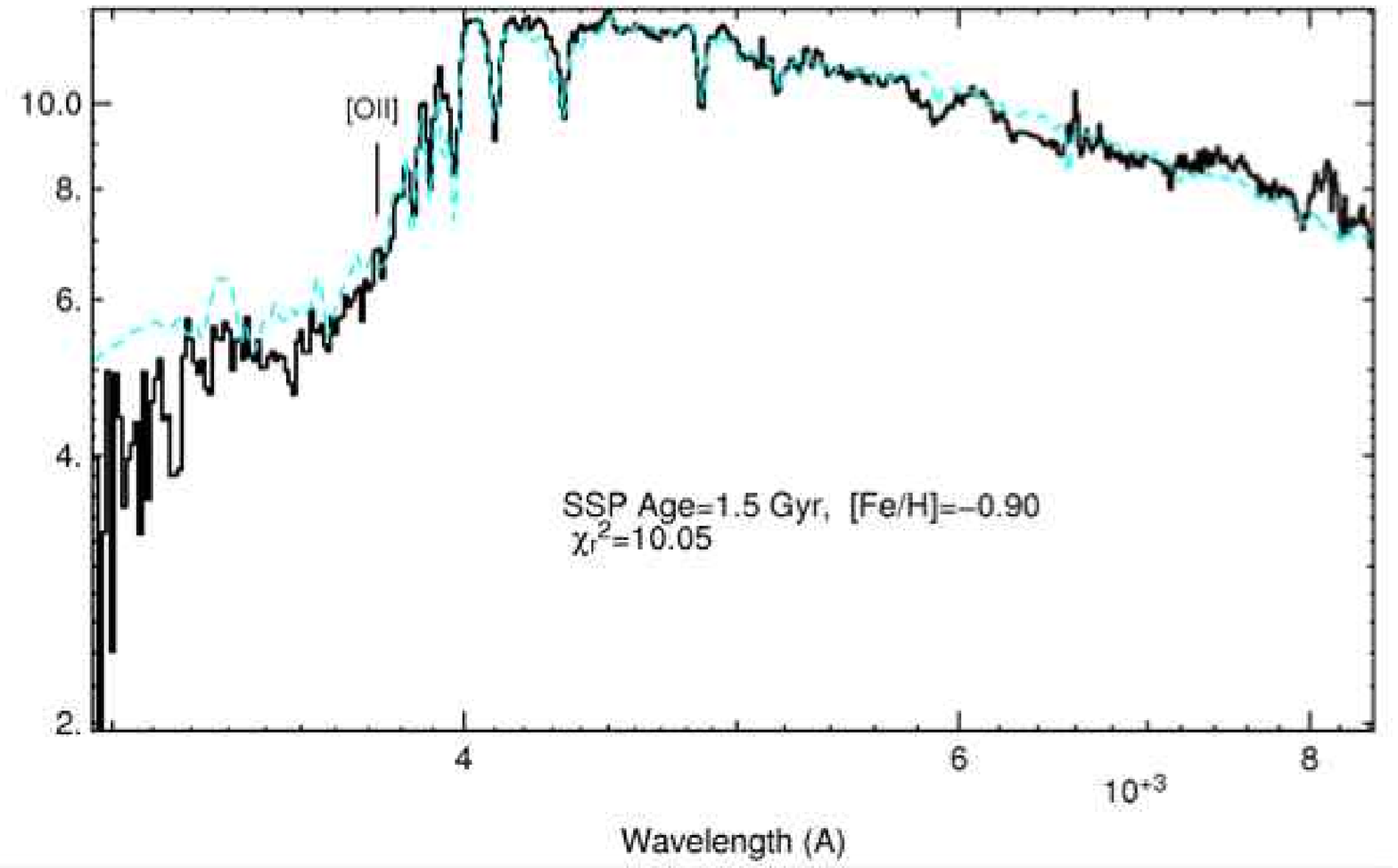}
\includegraphics[scale=0.40]{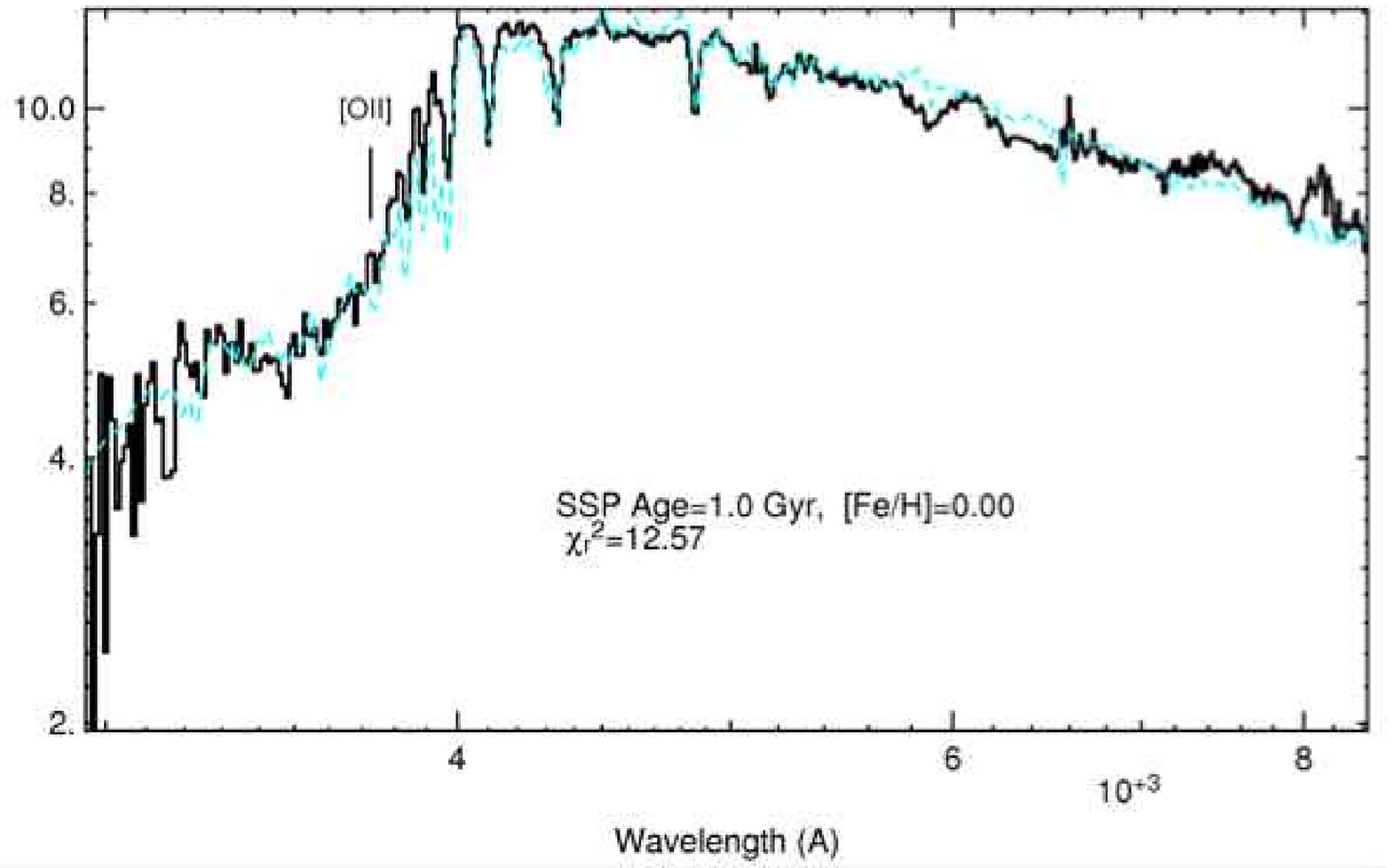}
\includegraphics[scale=0.40]{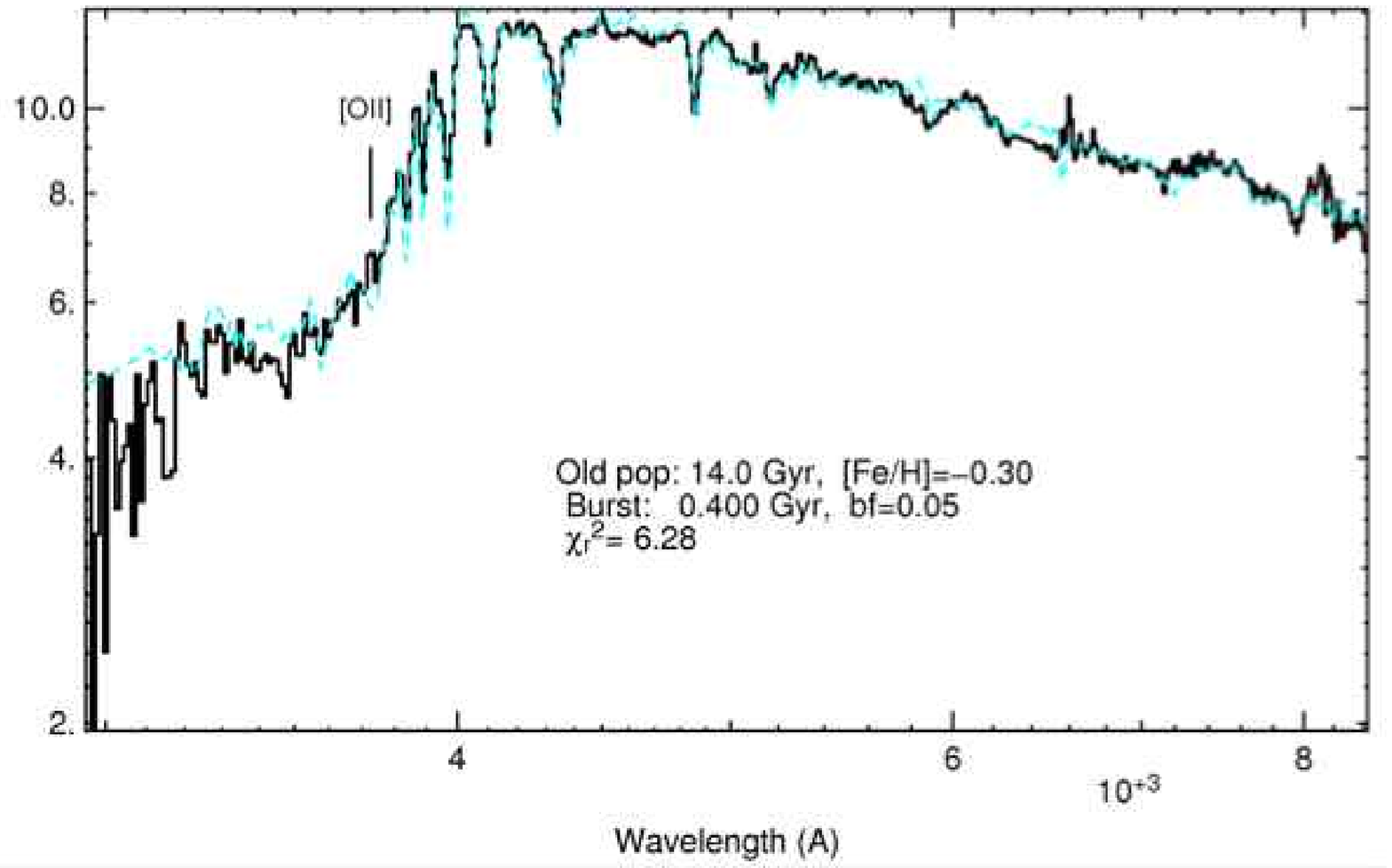}
\includegraphics[scale=0.40]{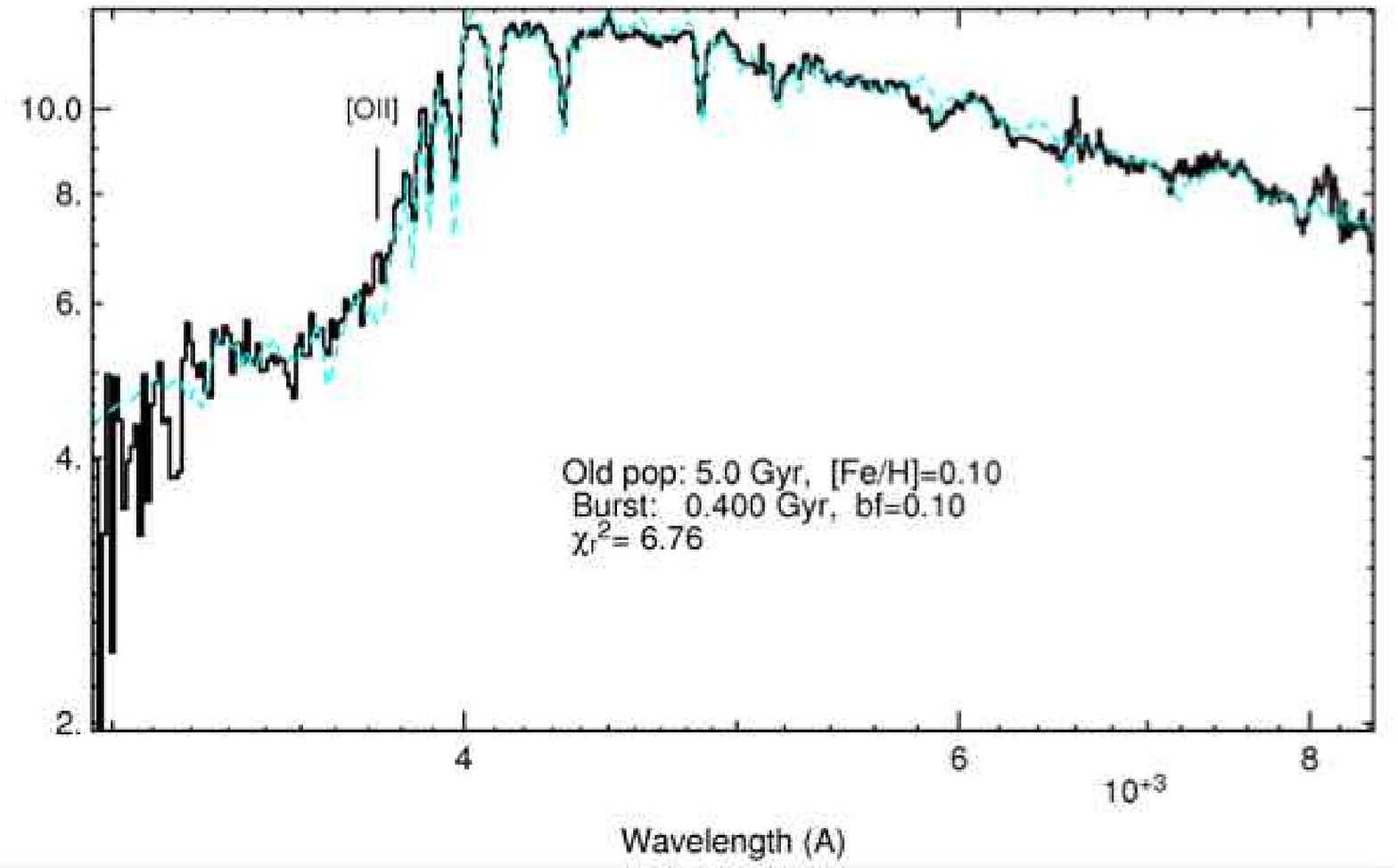}
\caption{
Stellar population model best-fits to the full spectrum of G515, in the rest-frame wavelength 
range $3200 - 8000$\AA.  The solid dark line is the spectrum and the dashed blue lines are 
the model fits.  The best-fit parameters of stellar population age, burst fraction, and metallicity,
as well as the reduced $\chi^2$ values for each fit, are given in each panel.  {\it upper left:} a simple starburst fit with metallicity as a free parameter.  {\it upper right:} simple starburst fit with metallicity fixed near solar.    {\it lower left:} a two-component fit with metallicity as a free parameter.  {\it lower right:} a two-component fit with metallicity fixed near solar.
} 
\label{starf}
\end{figure}

\clearpage
\begin{figure}[ht]
\epsscale{1.0} 
\includegraphics[scale=0.40]{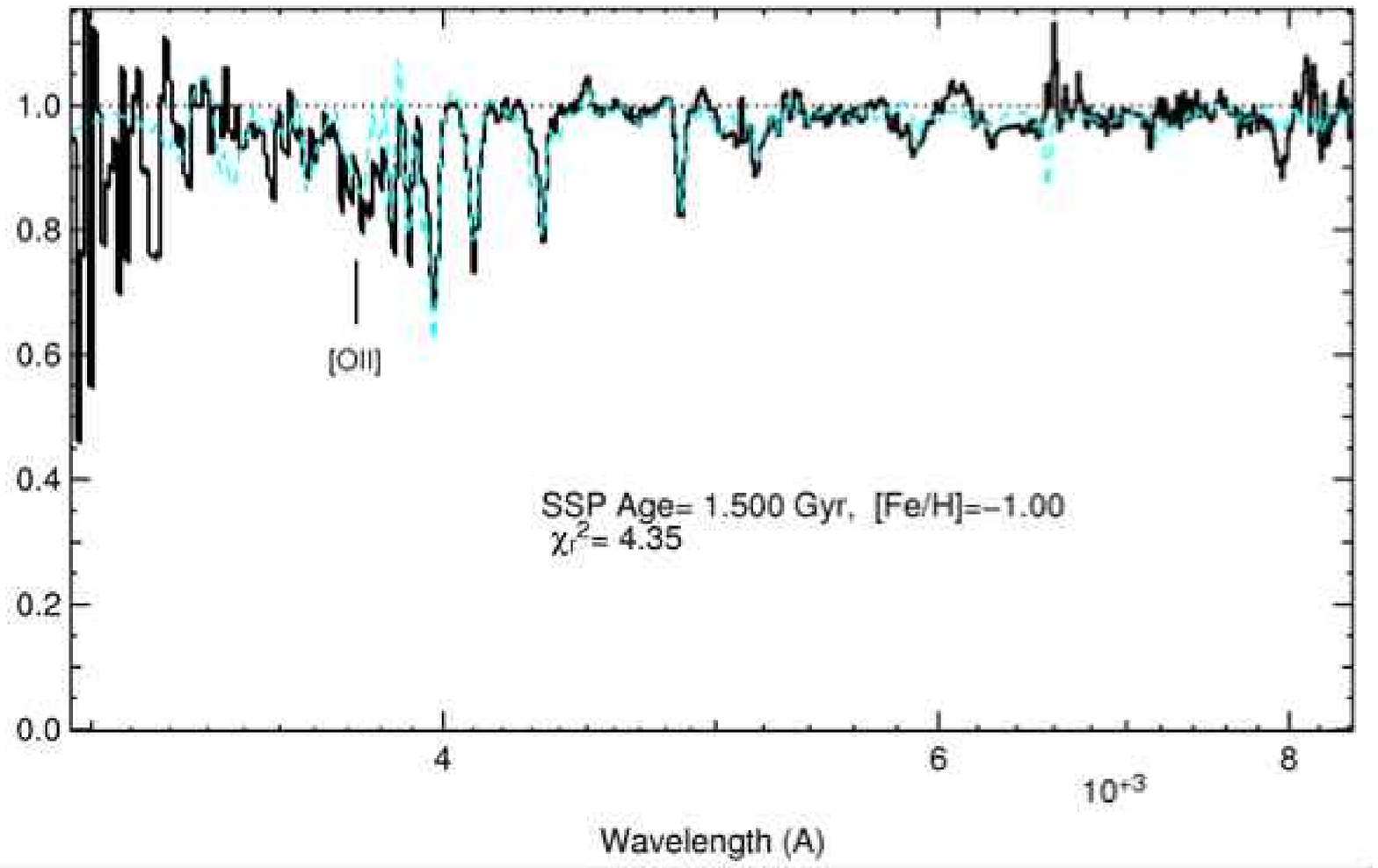}
\includegraphics[scale=0.40]{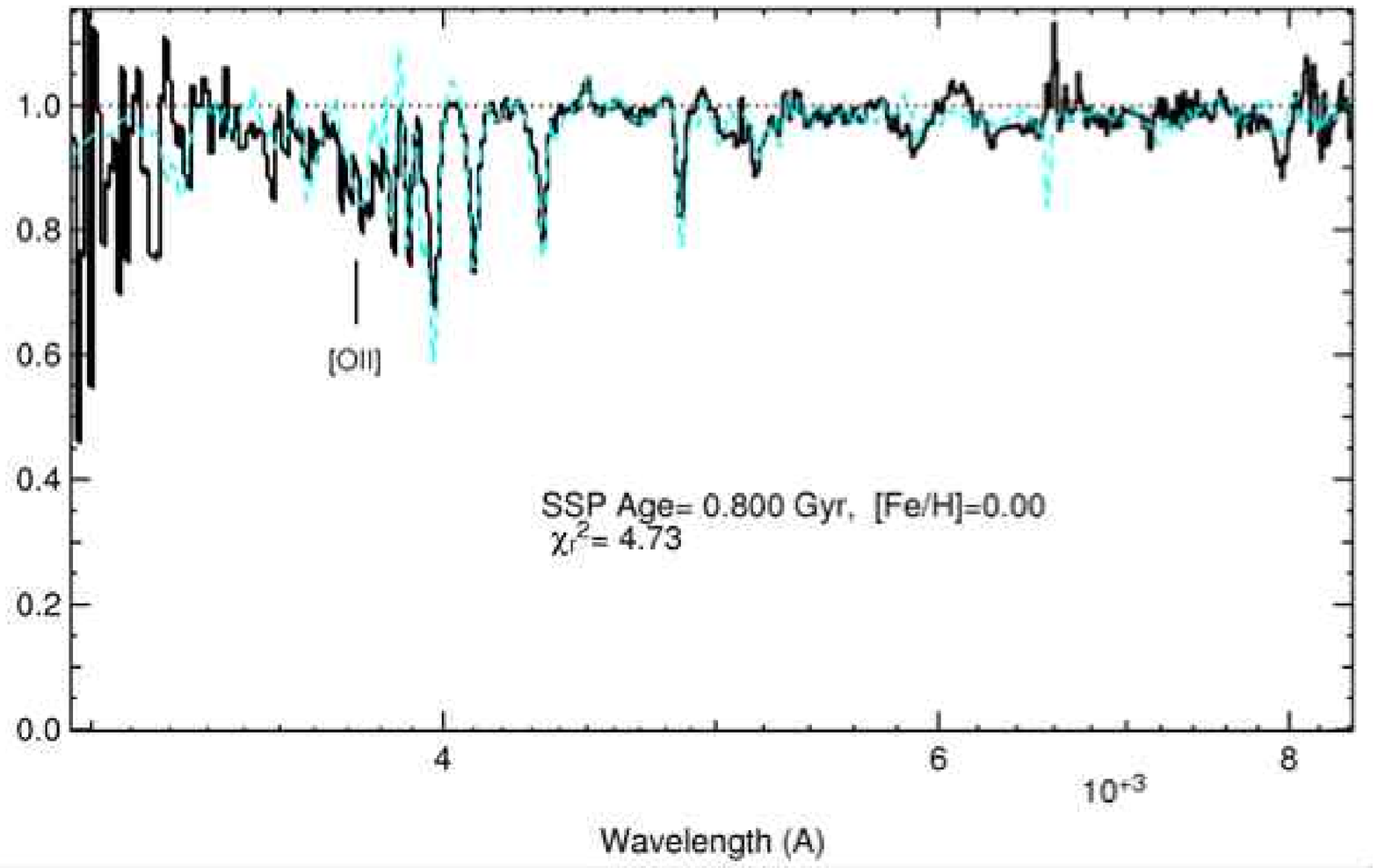}
\includegraphics[scale=0.40]{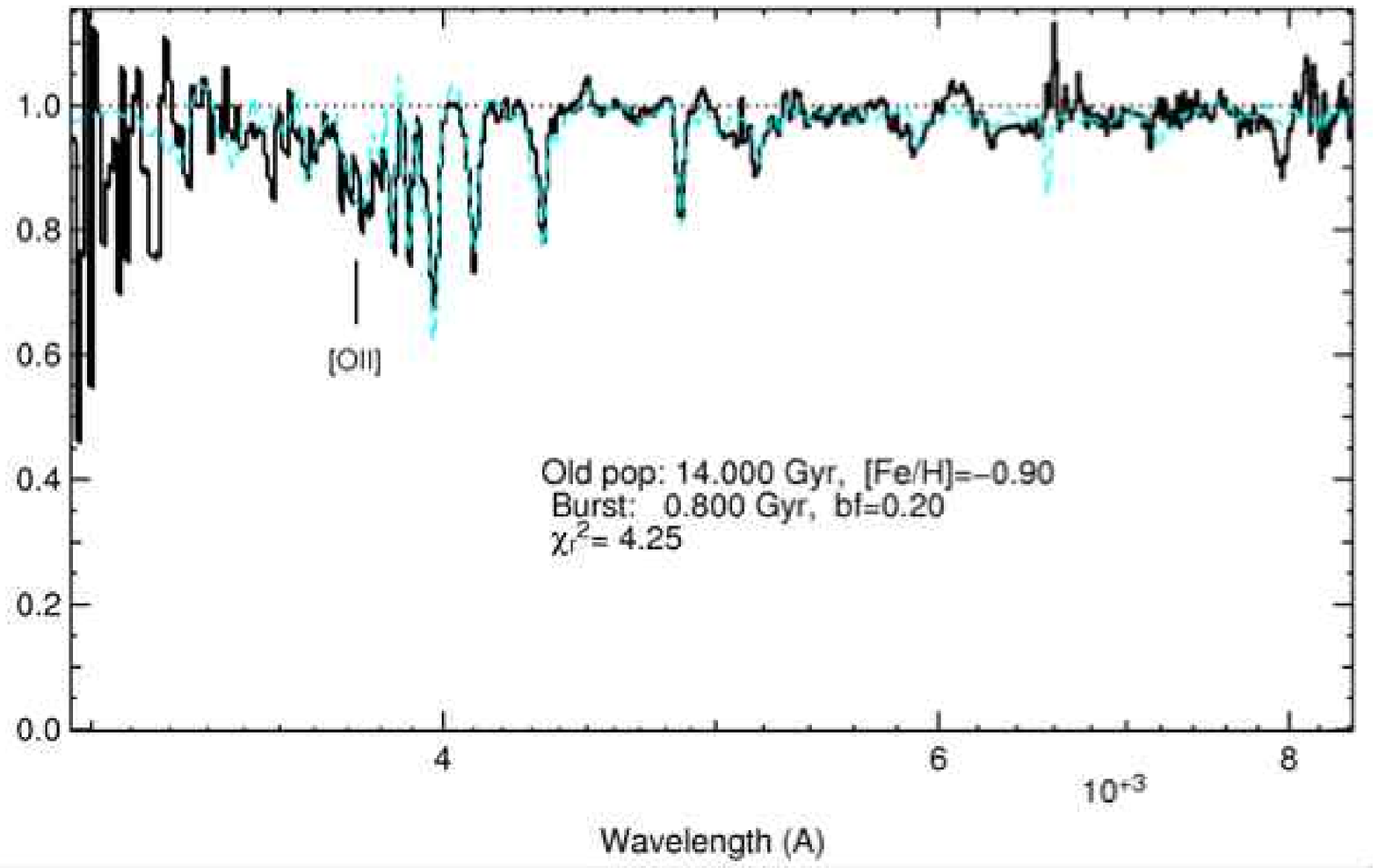}
\includegraphics[scale=0.40]{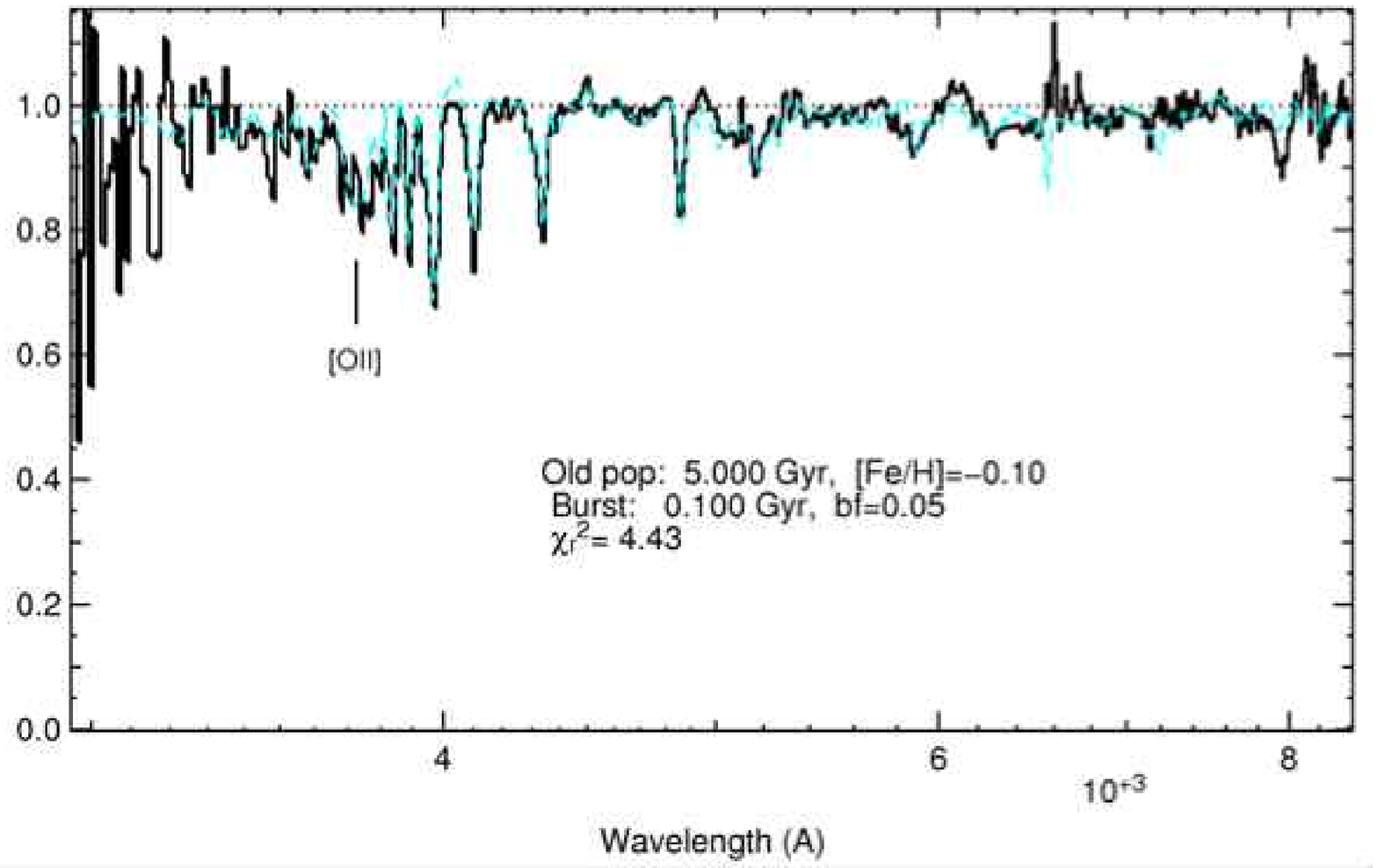}
\caption{
Stellar population model best-fits to the continuum-normalized spectrum of G515, in the rest-frame wavelength range $3200 - 8000$\AA.  The lines and panels are the same as in Figure \ref{starf}.
} 
\label{starc}
\end{figure}

 
 \end{document}